\documentclass{appolb}
\usepackage{epsfig}

\begin{document}
\title{Comparative analysis of large $N_c$ QCD and quark model approaches to baryons%
\thanks{Excited QCD 2009, February 8-14, Zakopane, Poland}%
}
\author{F. Buisseret\thanks{F.R.S.-FNRS Postdoctoral Researcher; E-mail: fabien.buisseret@umh.ac.be}, C. Semay\thanks{F.R.S.-FNRS Senior Research Associate; E-mail: claude.semay@umh.ac.be}
\address{University of Mons, Place du Parc 20, B-7000 Mons, Belgium}\\
\medskip
Fl. Stancu\thanks{E-mail: fstancu@ulg.ac.be}
\address{University of Li\`ege, Insitute of Physics B5, Sart Tilman, B-4000 Li\`ege, Belgium }
\and
N. Matagne\thanks{E-mail: Nicolas.Matagne@theo.physik.uni-giessen.de}
\address{Institut f\"{u}r Theoretische Physik, Universit\"{a}t Giessen, D-35392 Giessen, Germany}
}
\maketitle
\begin{abstract}
We show that a remarkable compatibility exists between the results of a potential
model with constituent quarks and the $1/N_c$ expansion mass formula for strange and
nonstrange baryon resonances. Such compatibility brings support to the basic
assumptions of relativistic quark models and sheds light on the physical content of
the model-independent large $N_c$ mass formula. Good agreement between both approaches is also found
for heavy baryons, made of one heavy and two light quarks, in the ground state band. 
\end{abstract}
\PACS{11.15.Pg, 12.39.Ki, 12.39.Pn, 14.20.-c}
  
 \section{Baryons in large $N_c$ QCD}\label{seclnc}
\subsection{Light baryons}
In large $N_c$ QCD, the gauge group is SU($N_c$) and a baryon is made of $N_c$ quarks.
The $1/N_c$ expansion is based on the discovery that, in the limit
$N_c \rightarrow \infty$, QCD possesses an exact contracted
SU(2$N_f$) symmetry where $N_f$ is the number
of flavors. This symmetry is approximate for finite $N_c$ so that corrections have to be added
in powers of $1/N_c$. When the SU($N_f$) symmetry is exact the mass operator $M$ has the general form 
\begin{equation}
\label{massoperator}
M = \sum_{i} c_i O_i.
\end{equation}
The coefficients $c_i$ encode the QCD dynamics and have to be determined from a fit to available data, while the operators $O_i$ are SU(2$N_f$) $\otimes$ SO(3) scalars of the form
\begin{equation}\label{OLFS}
O_i = \frac{1}{N^{n-1}_c} O^{(k)}_{\ell} \cdot O^{(k)}_{SF}.
\end{equation}
Here $O^{(k)}_{\ell}$ is a $k$-rank tensor in SO(3) and $O^{(k)}_{SF}$
a $k$-rank tensor in SU(2)-spin,
but invariant in SU($N_f$)-flavor. $n$
represents the minimum of gluon exchanges to generate the operator. In practical applications, terms of order $1/N_c^2$ are neglected. 

One obviously has to set $N_f=2$ for light nonstrange baryons. As an example, the ground state mass formula reads in this case $M = c_1 N_c \textbf{1}+ c_4 S^{\, 2}/N_c + \mathcal{O}\left(N_c^{-3} \right)$. Other terms like spin-orbit and isospin-dependent contributions appear in excited bands~\cite{Mata}. For light strange baryons ($N_f=3$), a general mass term of the form
\begin{equation}\label{break}
n_s ~\Delta M_s = \sum_{i=1} d_i B_i,
\end{equation}
has to be added to equation~(\ref{massoperator}) to account for SU(3)-flavor symmetry breaking. In the left hand side $n_s$ is the number of strange quarks and $\Delta M_s$ is the mass shift of every strange quark. The operators $B_i$ break SU(3)-flavor symmetry and the coefficients $d_i$ have to be fitted in a global fit of nonstrange and strange baryons.

The classification scheme used in the $1/N_c$ expansion for baryon resonances is based on the standard SU(6) classification as in a constituent quark model. Baryons are grouped into excitation bands $N= 0$, 1, 2,\dots, each band containing at least one SU(6) multiplet. The band number $N$ is the total number of excitation quanta in a harmonic oscillator picture. Note that the coefficients $c_i$ and $d_i$ depend on $N$. 

\subsection{Heavy quarks}

The approximate spin-flavor symmetry of baryons 
containing two light 
and one heavy 
quark is SU(6)$\times$ SU(2)$_c$ $\times$  SU(2)$_b$,
\emph{i.e.} there is a separate spin symmetry for each heavy flavor. For these baryons, an $1/m_Q$ expansion can be combined with the $1/N_c$ expansion, $m_Q$ being the heavy quark mass. In the case of an exact SU(3)-flavor symmetry, the mass operator reads
\begin{equation}\label{mlnc}
M=m_Q \textbf{1}+c_0\, N_c\, \textbf{1}+\frac{c_2}{N_c}\, J^2_{qq}
+\frac{c^{'}_0}{2 m_Q}  
\textbf{1}+\frac{c^{'}_2}{ 2 m_QN^2_c}J^2_{qq}+2 \frac{c^{''}_2}{N_c m_Q}\vec J_{qq}\cdot \vec J_Q,
\end{equation}
where $\vec J_{qq}$ ($\vec J_Q$) is the total spin of the light quark pair (of the heavy quark). The unknown coefficients have to be fitted to experimental data, but physical and dimensional arguments suggest to introduce a typical QCD energy scale $\Lambda$ and the relations
\begin{equation}
\label{largenpar}
c_0 =  \Lambda, \quad c_2 \sim \Lambda,\quad c^{'}_0  \sim   c^{'}_2\sim c^{''}_2  \sim  {\Lambda}^2.  
\end{equation}

The inclusion of SU(3)-flavor breaking leads to an expansion of the mass operator in the SU(3)-violating parameter $\epsilon \sim(m_s-m)\sim0.2-$0.3, where $m$ is the average mass of the $u$, $d$ quarks and where $m_s$ is the strange quark mass. Its value is measured in units of
the chiral symmetry breaking scale parameter $\Lambda_{\chi} \sim 1$ GeV.

\section{Quark model}
A baryon, viewed as a bound state of three quarks, can be described in a first approximation by the spinless Salpeter Hamiltonian 
\begin{equation}
	H=\sum^3_{i=1}\left[\sqrt{\vec p^{\, 2}_i+m^2_i}+
	\sigma |\vec{x}_{i}-\vec{R}|\right],
\end{equation}
where $m_i$ is the current quark mass and where $\sigma$ is the string tension. The confinement potential is a Y-junction in which the Toricelli point is replaced by the center of mass $\vec{R}$. It is also necessary to include some perturbative corrections, namely one-gluon exchange and quark self-energy mass terms, respectively 
\begin{equation}
\label{oge}
\Delta M_{\textrm{oge}}=-\frac{2}{3}\sum^3_{i<j=1}\left\langle \frac{\alpha_{s,ij}}{|
\vec x_i-\vec x_j|}\right\rangle,\quad  \Delta M_{\textrm{qse}}=-\frac{fa}{2\pi} \sum^3_{i=1} \frac{\eta(m_i/\delta)}{\mu_i}, 
\end{equation}
where $\alpha_{s,ij}$ is the strong coupling constant between the quarks $i$ and $j$ and $\mu_i=\left\langle \sqrt{\vec p^{\, 2}_i+m^2_i}\right\rangle$ is the kinetic energy of the quark $i$. The factors $3 \leq f\leq 4$ and ($1.0 \leq \delta \leq 1.3$)~GeV have been computed in lattice QCD. $\eta(x)$ is analytically known and can accurately be fitted by $\eta(x)\approx 1-\beta x^2$  with $\beta=2.85$ for $0 \le x \leq 0.3$ and by $\gamma/x^2$ with $\gamma=0.79$ for $1.0 \leq x \leq 6.0$. 

Within our model, we have $m_u=m_d=0$. In this case, 
using the auxiliary field technique, analytical mass formulas can be obtained for both light $qqq$ and heavy $qqQ$ baryons at order $\mathcal{O}(m_s^2)$ and $\mathcal{O}(1/m_Q)$. For light baryons one has $M_{qqq}=M_0 + n_s\, \Delta M_{0s}$ ($n_s=0,1,2,3$) with \cite{lnc1,lnc2}
\begin{eqnarray}
\label{M_qqq}
M_0 &=&6\mu_0-\frac{2 \pi\sigma \alpha_0}{6\sqrt{3}\mu_0}-\frac{f \sigma}{4\mu_0k_0}, \quad
\Delta M_{0s}=\frac{m^2_s}{\mu_0}\Theta(N).
\end{eqnarray}
We refer the reader to~\cite{lnc2} for the explicit expression of $\Theta(N)$. In these equations, $\mu_0=\sqrt{\pi\sigma(N+3)/18}$, and $\alpha_0=\alpha_{s,qq}$. $k$ is a corrective factor equal to $k_0=0.952$ ($k_1=0.930$) for $qqq$ ($qqQ$) baryons, resulting from the replacement of the Toricelli point by the center of mass. Moreover, $N$ is the baryon band number in a harmonic oscillator picture, just as the one which is used in large $N_c$ QCD as in section~\ref{seclnc}. This allows a direct comparison between both approaches. 

For heavy baryons one has $M_{qqQ}=m_Q+M_1 + n_s\, \Delta M_{1s} + \Delta M_{Q}$ ($n_s=0,1,2$), with ~\cite{lnc3}
\begin{eqnarray}
\label{M_qqQ}
M_1 &=&  4\mu_1 - \frac{2}{3}\left( \alpha_0\sqrt{\frac{k_1 \pi\sigma}{18k_0}} 
+ 2 \alpha_1\sqrt{\frac{ k_1 \pi\sigma}{3k_0(N+3)}}\right)-\frac{f \sigma}{6k_0\mu_1}, \nonumber \\
\Delta M_{1s}&=&\frac{m_s^2}{\mu_1}\bar\Theta(N),\quad  \Delta M_{Q}=\frac{k_1 \pi\sigma}{12k_0m_Q}\, K(N).
\end{eqnarray}
The interested reader will find the explicit expressions of $\bar\Theta(N)$ and $K(N)$ in Ref.~\cite{lnc3}.
Moreover, $\mu_1=\sqrt{k_1 \pi\sigma(N+3)/12k_0}$ and $\alpha_1=\alpha_{s,qQ}=0.7\alpha_0$. The band number $N$ corresponds this time to the relative quantum of excitation of the heavy quark and the light quark pair. The heavy quark--light diquark picture is favored since the quark pair tends to remain in its ground state~\cite{lnc3}.  

\section{Comparison of both approaches}
\subsection{Light baryons}
\begin{figure}[ht]
\includegraphics*[width=12.5cm]{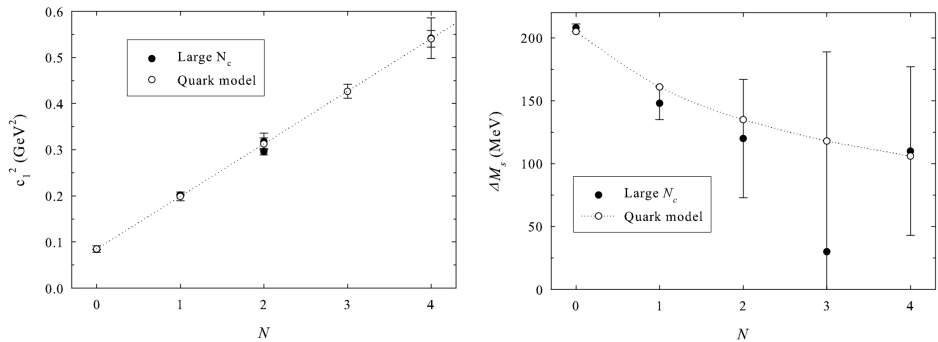}
\caption{Plot of $c^2_1$ (left) and $\Delta M_s$ (right) versus the band number $N$. The values computed in the $1/N_c$ expansion (full
circles) from a fit to
experimental data are compared with the quark model results with $\sigma=0.163$~GeV$^2$, $\alpha_0=0.4$, $f=3.5$, and $m_s=0.243$~GeV (empty circles and dotted line are given to guide the eyes).}
\label{Fig1}
\end{figure}

The coefficients $c_i$ appearing in the $1/N_c$ mass operator can be obtained 
from a fit to experimental data and compared with the quark model results. The dominant term $c_1\, N_c$ in the mass
formula~(\ref{massoperator}) contains the 
spin- and strangeness-independent mass contributions, which in a
quark model language represents the confinement and the
kinetic energy. So, for $N_c = 3$, we expect  
\begin{equation}\label{c1qm}
c^2_1=M^2_{0}/9.
\end{equation}
Figure~\ref{Fig1} (left panel) shows a comparison between the values of $c^2_1$
obtained in the $1/N_c$ expansion method and those derived from Eq.~(\ref{M_qqq}) for various values of $N$. It appears that the results of large $N_c$ QCD are entirely
compatible with the formula~(\ref{c1qm}) for standard values of the parameters.
The spin-dependent corrections between quarks $i$ and $j$ should be of order $O(1/\mu_i\mu_j)$. Therefore we expect both $c_2$ and $c_4$ to be proportional to $(N+3)^{-1}$: Such a behavior is consistent with the large $N_c$ results, where it is also observed that the spin-spin contribution ($c_4$) is much larger than the spin-orbit contribution ($c_2$)~\cite{lnc1}. 

The mass shift due to strange quarks is given in the
quark model by $\Delta M_{0s}$. A comparison of 
this term with its large $N_c$ counterpart
is given in Fig.~\ref{Fig1} (right panel), where we can see that
the quark model predictions are always
located within the error bars of the large $N_c$ results. In both approaches, one
predicts a
mass correction term due to SU(3)-flavor breaking which decreases with $N$.

\subsection{Heavy baryons}
The heavy quark masses $m_c$ and $m_b$ can be independently fitted to the experimental data in both the quark model and the $1/N_c$ frameworks~\cite{lnc3}. In large $N_c$ QCD one obtains $m_c=1315$~MeV and $m_b=4642$~MeV, while the quark model 
mass formula~(\ref{M_qqQ}) is compatible with the experimental data provided that $m_c=1252$~MeV and $m_b=4612$~MeV (the other parameters have been fitted to light baryons). Both approaches lead to quark masses that 
differ by less than 5\%. 

The other parameter involved in the large $N_c$ mass formula is $\Lambda$, which in the ground state band can be identified to the mass formula (\ref{M_qqQ}) as follows: $\Lambda=c_0=\frac{1}{3} \left.M_1\right|_{N=0}$. According to the large $N_c$ fits one has $c_0 = \Lambda \simeq 0.324$~GeV while the quark model gives 0.333~GeV,
which means a very good agreement for the QCD scale $\Lambda$. The terms of order $1/m_Q$ lead to the identity $c^{'}_0 = 2 m_Q \left.\Delta M_Q\right|_{N=0}$. The large $N_c$ parameter $\Lambda=0.324$~GeV gives $c^{'}_0\sim \Lambda^2 = 0.096$~GeV$^2$ and the quark model gives 0.091 GeV$^2$, which is again a good agreement. Finally, the SU(3)-flavor breaking term is proportional to 
$\epsilon\Lambda_\chi \sim m_s$. One should have $\epsilon\Lambda_\chi=\frac{2}{\sqrt 3}\left.\Delta M_{1s}\right|_{N=0}$ by definition of $\epsilon\Lambda_\chi$; indeed the large $N_c$ value $\epsilon\Lambda_\chi=0.206$~GeV and the quark model estimate
$0.170$~GeV also compare satisfactorily. We point out that, except for $m_c$ and $m_b$, all the model parameters 
are determined from theoretical arguments combined with phenomenology, or are 
fitted on light baryon masses. The comparison of our results with the 
$1/N_c$ expansion coefficients $c_0$, $c^{'}_0$ and $\epsilon\Lambda_\chi$ are 
independent of the $m_Q$ values. So we can say that this analysis is 
parameter free. 

An evaluation of the 
coefficients $c_2$, $c_2'$, and $c_2''$ through the computation of the 
spin-dependent effects is out of the scope of the present spin-independent formalism. But at the dominant order, the ratio $c_2''/c_2$ should be similar to 
$\mu_1=356$~MeV, which is roughly in agreement with 
equation~(\ref{largenpar}) stating that $c_2''/c_2 \sim \Lambda$. 

\section{Conclusion}
We have established a connection between the quark model and 
the combined $1/N_c$, $1/m_Q$ expansion both for light baryons and for heavy baryons containing 
a heavy quark. Our results bring reliable 
QCD-based support in favor of the constituent quark model assumptions and lead 
to a better insight into the coefficients $c_i$ encoding the QCD dynamics 
in the $1/N_c$ mass operator. As an outlook, we mention that a combined quark model -- $1/N_c$ expansion could lead to predictions for excited heavy baryons masses or even for ground state masses of baryons with two heavy quarks. This work is in progress. 

\end{document}